\expandafter\def\csname ver@fixltx2e.sty\endcsname{}
 \PassOptionsToPackage{pdfpagelabels=false}{hyperref} 
\documentclass[useAMS,usenatbib]{mnras}

\usepackage{aas_macros}
\usepackage{amsmath}
\usepackage{amssymb}
\usepackage{float}
\usepackage{graphicx}
\usepackage{lscape}
\usepackage{url}
\usepackage{natbib}
\usepackage{placeins}
\usepackage{color}
\usepackage{wasysym}
\usepackage{hyperref}
\usepackage{scrextend}
\usepackage[normalem]{ulem}

\newcommand{\revision}[1]{#1}%\textcolor{red}{#1}}

\newcommand{\hide}[1]{}

\newcommand*{\figuretitle}[1]{  \textbf{#1} \par\medskip}

\begin{document}

\title{Cooling and Instabilities in Colliding Radiative Flows with Toroidal Magnetic Fields}

\author[R.N. Markwick et al.]
{
R.N. Markwick$^{1}$,  A. Frank$^{1}$, 
E.G. Blackman$^{1}$, J. Carroll-Nellenback$^{1}$, 
\newauthor 
\ S.V. Lebedev$^{2}$,  D.R. Russell$^{2}$\thanks{Present Address: Technische Universitaet Muenchen, Forschungs Neutronenquelle Heinz Maier-Leibnitz, Lichtenbergstrasse1, D-85748 Garching, Germany}, J.W.D. Halliday$^{3,2}$, L.G. Suttle$^{2}$
, P.M. Hartigan$^{4}$ 
\\
$^{1}$Department of Physics and Astronomy, University of Rochester, Rochester, NY\\
$^{2}$Blackett Laboratory, Imperial College, London SW7 2BW, United Kingdom\\
$^{3}$Clarendon Laboratory, University of Oxford, Parks Road, OX1 3PU, United Kingdom\\
$^{4}$Department of Physics and Astronomy, Rice University, Houston, TX
}

\date{}

\pagerange{\pageref{firstpage}--\pageref{lastpage}}
\maketitle
\label{firstpage}

\begin{abstract}
We report on the results of a simulation based study of colliding magnetized plasma flows.  Our set-up mimics pulsed power laboratory astrophysical experiments but, with an appropriate frame change, are relevant to astrophysical jets with internal velocity variations.  We track the evolution of the interaction region where the two flows collide.  Cooling via radiative loses are included in the calculation.  We systematically vary plasma beta ($\beta_m$) in the flows, the strength of the cooling ($\Lambda_0$) and the exponent ($\alpha$) of temperature-dependence of the cooling function.  We find that for strong magnetic fields a counter-propagating jet called a "spine" is driven by pressure from shocked toroidal fields. The spines eventually become unstable and break apart.  We demonstrate how formation and evolution of the spines depends on initial flow parameters and provide a simple analytic model that captures the basic features of the flow.

\end{abstract}

\begin{keywords}
Herbig–Haro objects --  ISM: jets and outflows -- instabilities -- MHD  --  methods: numerical -- shock waves
\end{keywords}

%.

%\pagebreak

%.

%\pagebreak

\section{Introduction}

Shocks arise in hypersonic plasma flows 
%as a result of %collisions (self-interactions), 
when faster moving material 
collides with 
slower material ahead of it.  Radiative shocks, where energy is lost to optically thin photons, can arise in a wide variety of physical settings, such as protostellar jets \citep{protostellarJet, Frankea2014}, supernova explosions \citep{Supernova}, gamma ray bursts \citep{GRB}, active galactic nuclei \citep{AGN}, and  High Energy Density Plasma (HEDP). Hydrodynamic and MHD evolution in collision zones is expected to lead to high degrees of heterogenity or "clumpiness" \citep{Hansen17}. In the context of protostellar jets, for example, flow collisions correspond observationally to chains of knots \citep{Hartigan11}.

\begin{figure}
    \centering
        \includegraphics[width=\columnwidth]{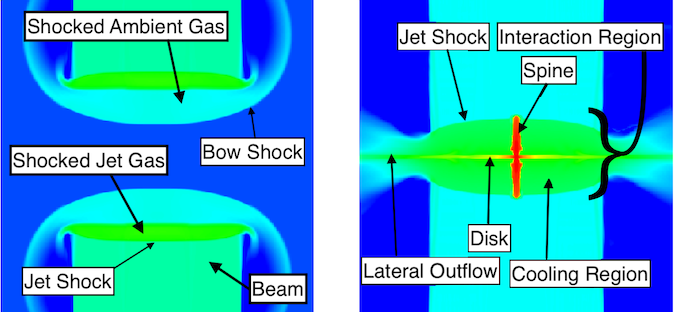}
    \caption{A diagram showing the structure of colliding MHD columns (jets).  The nature of structures such as the disk and spine are articulated in the text.  For comparison with a purely radiative hydrodynamic case see \citet{Markwick21}. }
    \label{fig:Schematic}
\end{figure}

Supersonic jets are structurally composed of a supersonic beam, a cocoon of shocked jet gas, a region of shocked ambient gas, and a bow shock \citep{blondin90}, as shown in figure \ref{fig:Schematic}.
When two supersonic flows collide, an \textit{interaction region} forms, bounded by a pair of \textit{jet shocks} also known as working surfaces. If radiative losses are significant, the interaction region will include a \textit{cooling region}, in which shocked gas cools from its initial post-shock temperature $T_s$. Further behind the shock, gas reaches its final post-cooling temperature $T_f$ and is deposited onto the \textit{cold slab}, where densities are highest.  Some shocked material gets ejected laterally by the high pressures throughout the interaction region, producing shocked lateral outflows (SLOs) \citep{Raga93, Markwick22}.

When magnetic fields are present, additional structures can form within the interaction region. One-dimensional simulations of pulsed jets conducted by \citet{Hartigan07} found that colliding magnetized blobs can essentially ‘bounce’ off of one another as magnetic pressure increases in cooling zones behind the shocks. \citet{Hansen15} investigated the lateral structure of this phenomenon with an axisymmetric model and found that while adiabatic shocks generate a net outward radial force in the jet column from high postshock pressures, the radial forces are inward for strongly-cooled isothermal shocks because the increased postshock magnetic pressure dominates the thermal pressure in that case. This inward motion produces a two-component structure: a disk and a spine. The disk spans the width of the interaction region in a manner similar to the cold slab in the hydrodynamic case \citep[as discussed in][]{Markwick22}, while the spine is the result of magnetic forces drawing cooled material towards the axis. The spine then expands along the z-axis and becomes a secondary jet that flows against the converging flows of the colliding columns. Figure \ref{fig:Schematic} contains a diagram highlighting the various features of hydrodynamic and MHD collisions.

Efforts have been made to study the MHD of plasma flow collisions in a laboratory setting, with differences in physical scale accounted for by the use of dimensionless parameters \citep{ryutov2000, ryutov2001, Falize2011}.  
Jets have been produced in the laboratory using both pulsed-power
\citep[e.g.][]{Lebedev05, Ciardi, gourdain10}
and laser-driven
\citep[e.g.][]{Laser1, Laser2, Laser3}
setups. 
Of particular relevence to this paper, \citet{suzukiVidal15} studied pairs of supersonic jets in collision. Among the results of this experiment was the emergence of small-scale structures within the interaction region.

In 
%first paper 
\cite{Markwick21}, we began a series of simulations  to study colliding radiative flows like those of \citet{suzukiVidal15}. The goal of this work has been to both recover the behavior seen in the experiments and understand the more general plasma dynamics of converging, magnetized radiative flows.  Our work began with a simplified model, featuring hydrodynamic flows and an analytic form of radiative cooling.
These simplifications allowed us to focus our attention on instabilities in the interaction region (specifically the cold slab).  In this way they provided a starting point for explaining the origin of small scale structures seen in the experiments. Long-term evolution of the simulations was found to be dominated by bending modes characteristic of the \textit{nonlinear thin shell instability}  \citep{Vishniac94}, which could be triggered either by sufficiently short cooling lengths or by oscillations resulting from the \textit{radiative shock instability} \citep{Langer81}. Meanwhile, we found no evidence for the \citet{field65} instability in these simulations.

%\bobbie{TODO Decide if paper 2 is even relevant to this one.}
% If not relevant, COMMENT this paragraph out instead of deleting so that it can be used for paper 4
In another paper \citep{Markwick22}, we continued our work by focusing on the larger-scale effects resulting from variation of flow parameters, namely density, velocity, and jet radius. While our simulations remained in the hydrodynamic limit, we did begin to use a detailed cooling function $\Lambda(n,T)$ which was computed to mimic the cooling behaviour of Aluminium (the material used in the laboratory experiments). When densities or velocities of the two flows are not identical, a net motion of the interaction region can result.  Meanwhile a difference in cross-sectional area (i.e. jet radius) results in a deflection of the angle at which shocked lateral outflow emerges from the interaction region.

%{\bf (Adam: Add short paragraph about second paper)}

In this paper, we continue our study of colliding radiative flows by examining the effects of magnetic fields. We limit initial magnetic fields to toroidal geometry. In order to better focus on instabilities, we return to the analytic cooling model used in \citet{Markwick21}. Cooling parameters $\Lambda_0$ and $\alpha$ are varied, as is the strength of the magnetic fields. Variation of both of these cooling parameters allows us to cover cases of both shorter and longer cooling lengths as well as the presence and absence of the radiative shock instability. 

This paper is organized as follows: 
In Section \ref{sec:meth} we discuss the model system and simulation parameters.
In section \ref{sec:result}, we present the results of the simulations. Section \ref{sec:discuss} will include a discussion of spine growth and instabilities of the interaction region. In sections \ref{sec:exp} and \ref{sec:conc}, we relate our findings to the laboratory and astrophysical settings, respectively.

\section{Methods and Model} \label{sec:meth}

The simulations in this study were conducted using AstroBEAR\footnote{https://astrobear.pas.rochester.edu/} \citep{cunningham09,carroll13}, which is a massively parallelized adaptive mesh refinement (AMR) code that includes a variety of multiphysics solvers, such as  radiative transport, self-gravity, heat conduction, and ionization dynamics.  Our study uses the magnetohydrodynamic solvers with an energy source term associated with the radiative cooling. 
Simulations were conducted in three dimensions in cartesian coordinates. Our governing equations are
\begin{subequations}
\begin{equation}
    \frac{\partial \rho}{\partial t} + \boldsymbol{\nabla} \cdot \rho \boldsymbol{v} = 0 
    \label{eq:Eu1}
\end{equation}
\begin{equation}
    \frac{\partial \rho \boldsymbol{v}}{\partial t} + \boldsymbol{\nabla} \cdot \left ( \rho \boldsymbol{v} \otimes \boldsymbol{v}- \frac{1}{4\pi}\boldsymbol{B}\otimes\boldsymbol{B} + p\boldsymbol{I}\right ) = 0
    \label{eq:Eu2}
\end{equation}
\begin{equation} 
    \frac{\partial E}{\partial t} + \boldsymbol{\nabla} \cdot \left((E + p) \boldsymbol{v} - \frac{(\boldsymbol{v}\cdot\boldsymbol{B})}{4\pi} \boldsymbol{B}\right) =-n^2\Lambda(T)
    \label{eq:Eu3}
\end{equation}
\begin{equation} 
    \frac{\partial \boldsymbol{B}}{\partial t} = \boldsymbol{\nabla}\times(\boldsymbol{v}\times\boldsymbol{ B})
    \label{eq:Eu4}
\end{equation}
\end{subequations}

where $\rho$ is the mass density, $n$ is the number density of nuclei, $\boldsymbol{v}$ is the fluid velocity, $p = p_\text{th}+\frac{1}{8\pi}B^2$ is the combined thermal ($p_\text{th}$) and magnetic pressure, $\boldsymbol{I}$ is the identity matrix, and $E = \frac{1}{\gamma - 1} p_\text{th} + \frac{1}{2}\rho v^2 + \frac{1}{8\pi}B^2$ is the combined internal and bulk kinetic energies. In all runs an average particle mass of 1 amu was used. $n^2\Lambda(n,T)$ is the cooling funciton, which gives the rate of radiative energy loss per unit time per unit volume. It may be worth noting that for this work we ignore magnetic resistivity; this will be considered in a future work.

As in \citet{Markwick21}, we use a power-law cooling function\footnote{Note that parameters $\alpha$ and $\beta$ from \citet{Markwick21} have been relabeled as $\Lambda_0$ and $\alpha$ to avoid conflicting with the plasma-beta} of the form $\Lambda(T) = \Lambda_0\left(\frac{T}{T_0}\right)^\alpha$. Cooling is applied only at temperatures above the floor temperature of the simulation in order to safeguard against runaway cooling. Since realistic cooling curves tend to vanish at low temperatures, this behaviour is justifiable on physical grounds.
Reference temperature $T_0$ was fixed at $2.25\times10^4$ K (1.94 eV) across all runs.

\revision{The simulations were conducted in a space of 64 by 64 by 64 computational units, with one computational unit corresponding to 0.1 mm. Within a radius of 16 computational units (centred on the jet axis), up to three levels of refinement were permitted for a maximum resolution of 0.0125 mm; outside this radius, refinement was limited to two levels for a maximum resolution of 0.025 mm.}
Extrapolated boundary conditions (specifically, Neumann boundary conditions with a derivative of zero) were used in all directions. 

Our simulations feature collisions of two cylindrical jets driven from the top and bottom z-boundaries with speed $v_i$ = 70 km s$^{-1}$. \revision{Jets were initialised as cylindrical regions with height 0.5 computational units, in which the density and velocity were held constant. }
In an effort to match the experiments of \citet{suzukiVidal15}, the jets were of density $3\times 10^{18}$ particles per cm$^3$, while the ambient medium was set to a density of $5\times10^{17}$ particles per cm$^3$ and a temperature of 4320 K.

\begin{figure}
    \centering
    \includegraphics[width=\columnwidth]{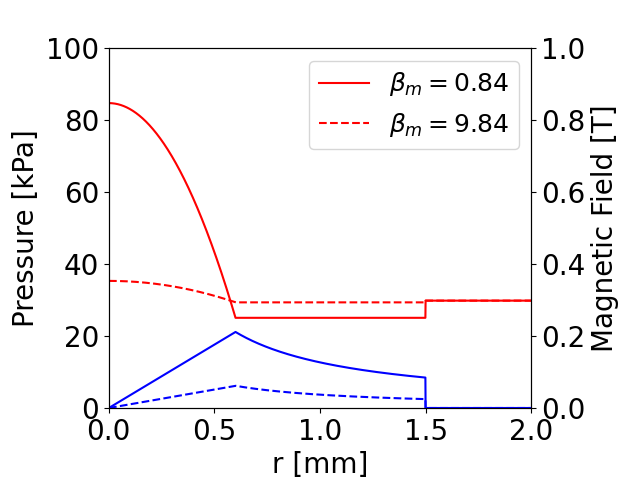}
    \caption{Thermal Pressure (red) and magnetic fields (Blue) in the initial (pre-collision) jets}
    \label{fig:initial}
\end{figure}

Initial conditions are such that hydromagnetic pressure equilibirum is established across the jet. To achieve this, we used a profile previously used by \citet{Hansen15} and others \citep[e.g.][]{Lind89, Frank98}.
The initial toroidal magnetic field and pressure inside the jet are given by
\begin{equation} \label{eq:B_initial}
B(r) = B_m \times \begin{cases}
 \frac{r}{R_m}  & 0 \leq r < R_m\\
\frac{R_m}{r}  & R_m \leq r < R_j\\
0 & R_j \leq r
\end{cases}
\end{equation} 
\begin{equation} \label{eq:p_initial}
 p(r) = \frac{B_m^2}{8\pi} \times\begin{cases}
\beta_m +2\left(\frac{R_m^2-r^2}{R_m^2}  \right) & 0 \leq r < R_m\\
\beta_m& R_m \leq r < R_j\\
\beta_m + \frac{R_m^2}{R_j^2}& R_j \leq r.
\end{cases}
\end{equation}
Here $R_m = 0.6$ mm is the radius at which the magnetic field is maximized at $B = B_m$, $R_j = 1.5$ mm is the jet radius, and $\beta_m = \frac{8\pi p(R_m)}{B_m^2}$ is the plasma-beta.  $B_m$ is a  parameter defining the maximum magnetic field, computed such that $p(R_j)$ is equal to the thermal pressure of the ambient medium. In the hydrodynamic limit, this gives a jet temperature of 710 K, though jet temperatures can be as low as 596 K in the $\beta_m = 0.84$ case. \revision{Figure \ref{fig:initial} shows these initial conditions.}

%p_amb = 298220 erg/cm3
%p(Rm) = p_amb * beta_m/(beta_m + (rm/rj)^2) = (0.84, 0.984, 1) * 298220
%c_s = sqrt(5/3 * p(Rm) / rho_i) = (0.9165, 0.9920, 1) * 3.158 km/s
%M = v_i/c_s = (1.091, 1.008, 1) * 22.166
%a = c_s * sqrt(2 / (gamma * beta_m) ) = (1, 0.3162, 0.00316) * 3.459 km/s
%sqrt(a^2+c_s^2) = (4.510, 3.318,3.158)
%M_ms =v_i / sqrt(a^2+c_s^2) = 15.52, 21.10, 22.17
 \begin{table}
     \centering
      \begin{tabular}{|c|c||c|c|c||c|c|}\hline
    Run & Figures & $\Lambda_0 / \Lambda_\text{ref}$& $\alpha$ & $\beta_m$ & $M$ & $M_{ms}$  \\ \hline\hline
1 & \ref{fig:run1}, \ref{fig:runs_1_3}, \ref{fig:runs_1_5}  & $2.0$ & +1 & 0.84 & 24.1 & 15.5\\\hline
2& \ref{fig:run2}  & $1.0$  & -1 & 0.84 & 24.2 & 15.5\\\hline
 3& \ref{fig:runs_1_3}  & $2.0$ & +1 & 9.84 & 22.3  & 21.1 \\\hline
4&\ref{fig:runs_4}   & $1.0$ & -1 & 9.84 & 22.3 & 21.1\\\hline
5& \ref{fig:runs_1_5}   & $20.0$ & +1 & 0.84 & 24.2 & 15.5\\\hline
6& \ref{fig:run_6_osc}  & $10.0$ & -1 & 0.84 & 24.2 & 15.5\\\hline
7& \ref{fig:runs_3_7}  & $20.0$  & +1 & 9.84 & 22.3  & 21.1\\\hline
8& \ref{fig:run8}  & $10.0$ & -1 & 9.84 & 22.34  & 21.1 \\\hline
9& \ref{fig:runs_9_10}  & $20.0$  & +1 & $10^5$ & 22.2& 22.2\\\hline
10& \ref{fig:runs_9_10}   & $10.0$ & -1 & $10^5$ & 22.2& 22.2 \\\hline
 \end{tabular}
     \caption{A summary of parameters varied between simulations, namely cooling strength $\Lambda_0$ (expressed in terms of $\Lambda_\text{ref} = 10^{-23}$  erg cm$^3$ s$^{-1}$), 
     cooling power $\alpha$, and plasma parameter $\beta = \frac{p(R_m)}{B_m^2/(8\pi)}$. 
     We also note the expected sonic $\left(M = \frac{v_i}{\sqrt{\gamma p(R_m)/\rho_i}}\right)$ and magnetosonic $\left(M_{ms} = M\sqrt{\frac{\gamma\beta_m}{\gamma\beta_m+2}}\right)$ mach numbers, computed post-collision using $\gamma = \frac53$. 
     }
\label{table:3DRuns}
\end{table}

Our study varies three parameters: one which describe magnetic fields and two which describe cooling. All parameters are summarised in table \ref{table:3DRuns}. First, we vary the plasma-beta 
%\eric{shouldn't  $p_{amb}$ replace $p_{th}$ for consistency below? But another point is that this definition of beta doesn't actually apply at any specific location as it uses the field at $R_M$ but pressure at exterior. It seems we should define a beta where the numerator and denominator are at the same position.  Natural position is $R_M$ otherwise its not a directly physically meaningful quantity and inconsistent with standard definitions which apply at a fixed position.}\bobbie{Good catch on the subscript. The reason it is defined that way is that this is how \citet{Hansen17} defined it, and I'm using that code. Could redefine and redo the calculations if desired - defining $\beta_m$ as value at $R_m$ means that we used 0.84 and 9.84 instead of 1 and 10.}\eric{I think it makes sense to correct their definition and set new standard :)}\bobbie{The conversion is $\beta_m = \beta_\text{old} - \frac{R_m^2}{R_j^2}$. Working on fixing throughout}
between 0.84 and 9.84. We also include two cases of $\beta_m = 10^5$, which corresponds to the hydrodynamic limit. The magnetic fields of the two jets are aligned in the same direction; this is done to correspond with both \citet{suzukiVidal15} as well as protostellor jets, as well as to limit the need to consider the effects of reconnection.

We selected our cooling parameters in a way which corresponds to different hydrodynamic instabilities as studied in \citet{Markwick21}. 
First, cooling power $\alpha$ is varied between $\alpha = -1$ and $\alpha = +1$. Based on results from  \citet{stricklandBlondin95} and \citet{Markwick21}, the former case is likely to exhibit the radiative shock instability \citep{Langer81}, while the latter is unlikely to do so.
In addition to the power law exponent, we also vary cooling strength $\Lambda_0$ between low and high strengths.
In \citet{Markwick21}, we found that momentum conservation for a $\gamma=\frac53$ gas gives a cooling length of approximately
\begin{equation} d_\text{cool}  = \frac{v_s}{4}\frac{5k T_s}{n_s\Lambda(T_s)} \left(\frac{1-\left(\frac{T_f}{T_s}\right)^{3-\alpha}}{3-\alpha}\right)      \label{eq:dcool} \end{equation}.
  We adjust the values of $\Lambda_0$ with $\alpha$ such that the cooling length remains fixed: Our `low' strengths are defined as $\Lambda_0 = 1.0 \times 10^{-23}$ erg
cm$^3$ s$^{-1}$ for $\alpha = -1$ and $\Lambda_0 = 2.0 \times 10^{-23}$ erg
cm$^3$ s$^{-1}$ for $\alpha = +1$, while the `high' strengths are equal to ten times the corresponding low strength.

\section{Results} \label{sec:result}

\subsection{Weak Cooling}
We begin our results with the four cases in which $\Lambda_0$ is of order $10^{-23}$ erg cm$^3$ s$^{-1}$. In all four of these cases we observe prominent formation of the spine, though specific details vary with both the magnetic field strength (characterized by $\beta_m$) and cooling behaviour (determined by exponent $\alpha$) are varied.  

\begin{figure}
    \centering
    \figuretitle{Low $\beta_m$, low $\Lambda_0$, $\alpha=+1$}
        \includegraphics[width=\columnwidth]{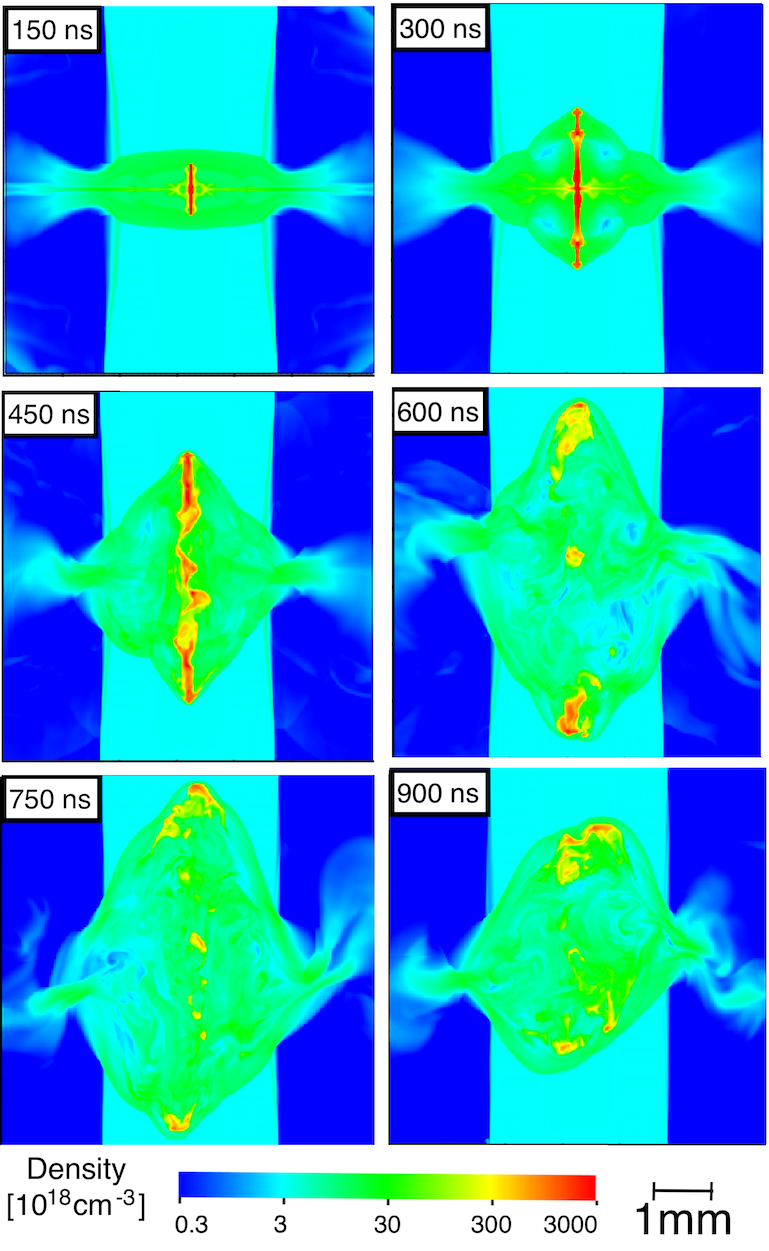}
    \caption{Run 1 from table \ref{table:3DRuns}, which features a relatively strong magnetic field, along with weak cooling which weakens as temperature decreases. A prominent feature is the formation of a spine, driven by magnetic hoop stresses ($\nabla B_\phi^2$), which eventually collapses as a result of an instability.  In this figure and all figures within seciton \ref{sec:result}, images are taken as midplane slices of density, and tick marks are spaced at 1mm intervals.  }
    \label{fig:run1}
\end{figure}

\begin{figure}
    \centering
    %\figuretitle{Spine Growth}
        \includegraphics[width=\columnwidth]{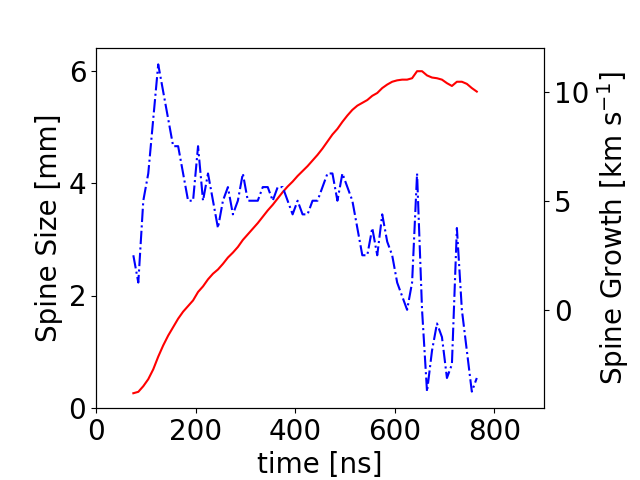}
    \caption{The size (red) and growth speed (blue) of the spine in Run 1 (see table \ref{table:3DRuns} or figure \ref{fig:run1}). After an initial surge, the growth speed remains roguhly constant until around 500 ns. After this time, the spine growth decelerates at an approximately constant rate. }
    \label{fig:run1_time}
\end{figure}

We begin with the case of strong magnetic field ($\beta_m = 0.84$) and a positive cooling exponent ($\alpha = +1.0$). This run can be seen in figure \ref{fig:run1} which shows density slices through the 3-D simulations. At the edges of the jet we see material ejected laterally into the surrounding medium. Such shocked lateral outflows (SLO) where seen in the hydrodynamic simulations \cite{Markwick21,Markwick22}.
These outflows are driven by the difference between high pressure within the interaction region and the lower pressure of the ambient medium. The initial lateral pressure discontinuity smooths into a gradient of thermal pressure spanning the outer portion ($r>R_m$) of the interaction region\footnote{Prior to any disturbance caused by the SLO or by pressure gradients of the inner region, the pressure within the outer region ($R_m< r < R_j$) of the interaction region is constant, and the net magnetic force in this region is zero owing to the variation of $B_\phi$ as $\frac{1}{r}$.}.

Almost immediately after the collision between the two counter-flowing jets, magnetic forces in regions of the cold slab closer to the axis become stronger than thermal pressure gradients in those regions. The dominance of magnetic forces in the inner region causes material to be drawn towards the axis, increasing the density \citep{Hansen15, Raga08}. 
Soon afterwards a spine - an axial bipolar flow propagating back into both jets - begins to emerge along the axis. The spine quickly grows to a length comparable to the diameter of the counter propagating jets. 
As the spine propagates, it pushes through jet material, deforming it. This back-reaction  modifies the structure of the interaction region and the two ``working surfaces" which bound it. \revision{After an initial surge and up through around 500 ns, the growth speed of the spine is approximately constant, as seen in figure \ref{fig:run1_time}.}

At later times, the spine becomes subject to what appears to be a form of a kink instability.  \revision{Once this occurs,} the disrupted spine is unable to support itself against the ram pressure of the jets and begins to \revision{decelerate back towards the centre of the interaction region.  This can be seen in figure \ref{fig:run1_time}, in which the growth speed of the spine undergoes approximately constant deceleration until it eventually begins to collapse inwards. In the process of collapse,} the entire interaction region is strongly distorted.

  The creation of the spine leads to a significant change in the evolution of the interaction region compared with the hydrodynamic cases \citep{Markwick21}, in which instabilities of the cooling region and the disk dominate long-term evolution of the interaction system. 

\begin{figure}
    \centering
    \figuretitle{Low $\beta_m$, low $\Lambda_0$, $\alpha=-1$}
        \includegraphics[width=\columnwidth]{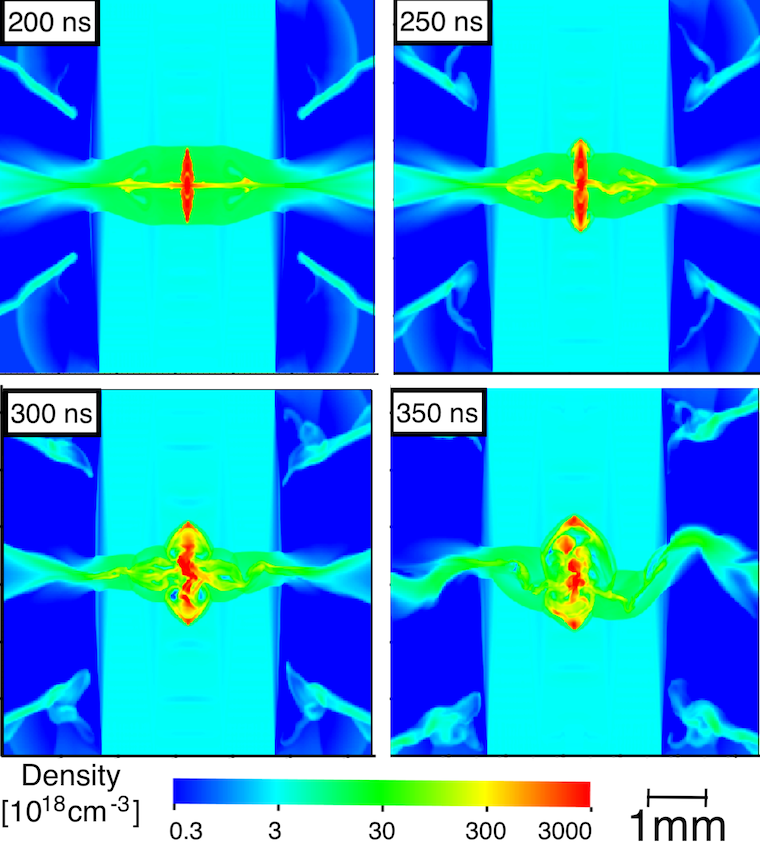}
    \caption{Run 2  from table \ref{table:3DRuns}, which features a relatively strong magnetic field along with weak cooling. Unlike the previous case, cooling strengthens as temperature decreases. Additional instabilities result in an earlier collapse of the spine. }
    \label{fig:run2}
\end{figure}

Next we examine Run 2 which is similar to the Run 1, but uses a different power law exponent. Run 1 had $\alpha = +1.0$, indicating that cooling is stronger at high temperatures and weakens as the material cools. Run 2 uses $\alpha = -1.0$, indicating that cooling is weaker at high temperatures but strengthens as the material cools. The initial postshock cooling strength is adjusted so that both runs feature similar cooling lengths, as can be seen by comparing figures \ref{fig:run1} and \ref{fig:run2}. 

Unlike in the $\alpha=+1.0$ case, the cold slab in the $\alpha=-1.0$ retains a noticeable amount of cooled material which we refer to as the disk. 
This is likely the result of the nature of density gradients within the cooling region: for the $\alpha=+1.0$ case, the gradient is appreciable throughout the cooling region; for cases with $\alpha<0$, since the cooling strengthens at lower temperatures there is a sharper transition to the cold slab (see figure 3 of \citet{Markwick21}).

After some time, the cold slab experiences significant perturbations. Oscillations drive cold material into the cooling regions above and below. As these perturbations grow they lead to interactions with the spine, which excites further instabilities.  We note that Run 2's perturbed cold slab leads to the kink occurring in the spine long before it was seen in Run 1.

\begin{figure}
    \centering
    \figuretitle{Low (left) vs high (right) $\beta_m$. Low $\Lambda_0$, $\alpha=+1$}
        \includegraphics[width=\columnwidth]{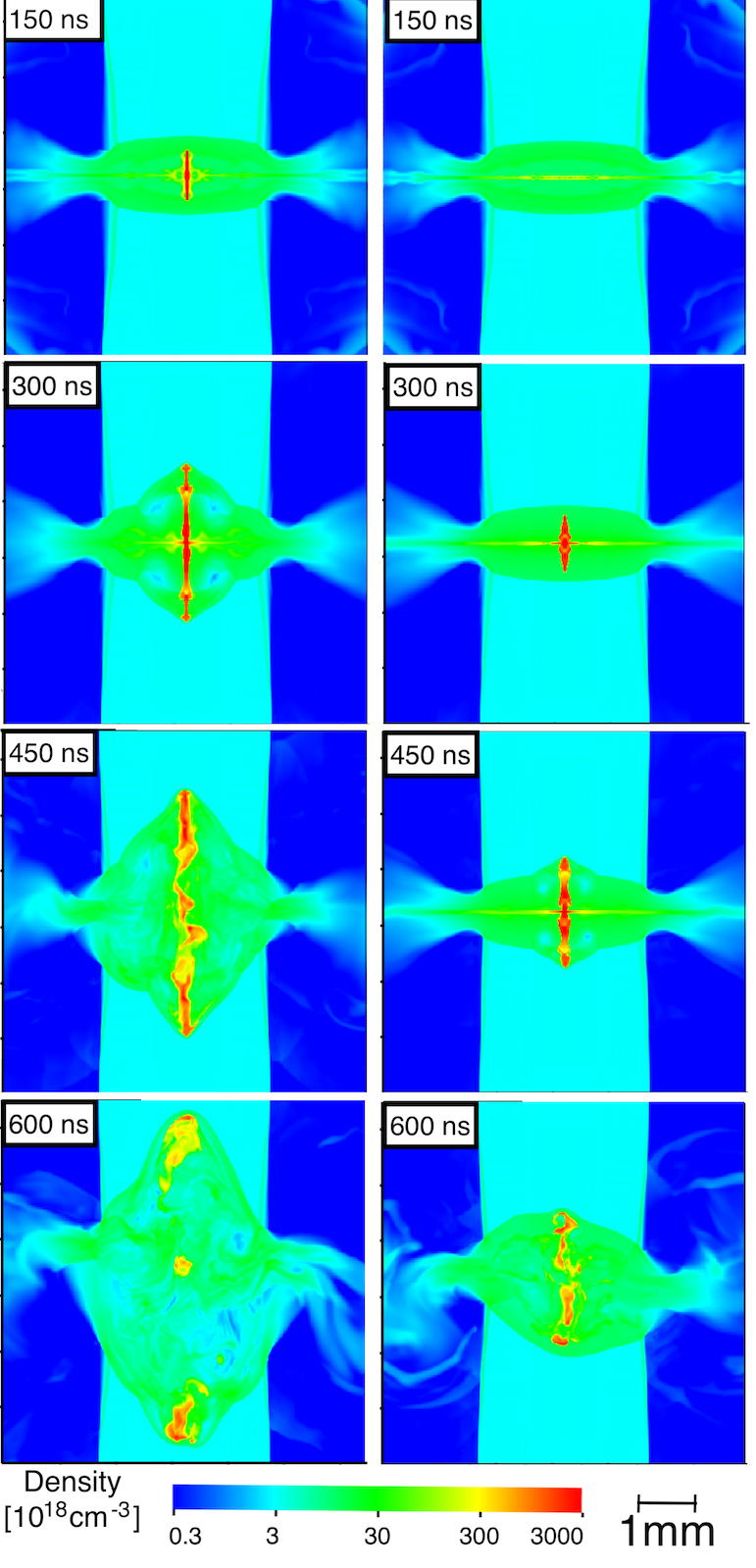}
    \caption{Runs 1 (left) and 3 (right) from table \ref{table:3DRuns}. Both cases feature identical cooling (weak cooling which weakens as temperature decreases), but run 3 has a weaker magnetic field. As a result, the spine forms more slowly and doesn't grow as large.  }
    \label{fig:runs_1_3}
\end{figure}

Our next case, Run 3, is shown on the right side of figure \ref{fig:runs_1_3}.  Run 3 features identical cooling to Run 1 (low $\Lambda_0$, $\alpha=+1.0$), but has a weaker magnetic field ($\beta_m = 9.84$).

The most significant differences in the results of Runs 1 and 3 (which in setup differ only in the value of $\beta_m$) are found in the spine. To understand these differences, however, it is first helpful to focus on the evolution of the disk. In the $\beta_m = 9.84$ case the disk more prominent than in its $\beta_m = 0.84$ counterpart, though it is not as prominent as it is in the $\alpha=-1.0$ (even-numbered runs) or high $\Lambda_0$ (Runs 5-10) cases. This is likely a result of magnetically-driven collimation forces being weaker which allows more of the cooled material to remain in the disk rather than being drawn into the spine. Since the disks is partly responsible ``feeding" the spine, such differences should be expected to effect the latter's evolution.

These expectations are born out in Run 3 as the initial formation of the spine takes nearly twice as long to begin as in Run 1. After spine formation begins subsequent growth is also significantly slower in the $\beta_m=9.84$ case (Run 3) than its $\beta_m=0.84$ counterpart (Run 1). This is likely the result of weaker magnetic fields driving the collimation process, especially in cases such as these where a limited amount of cold slab material is available for collimation. As with the $\beta_m = 0.84$ case, the spine is eventually disrupted by an effect which appears to be similar to the kink instability. This disruption begins at a similar duration after the initial spine growth begins, which results in a shorter spine length at the time of collapse. 

\begin{figure}
    \centering
        \figuretitle{High $\beta_m$, low $\Lambda_0$, $\alpha=-1$}
        \includegraphics[width=\columnwidth]{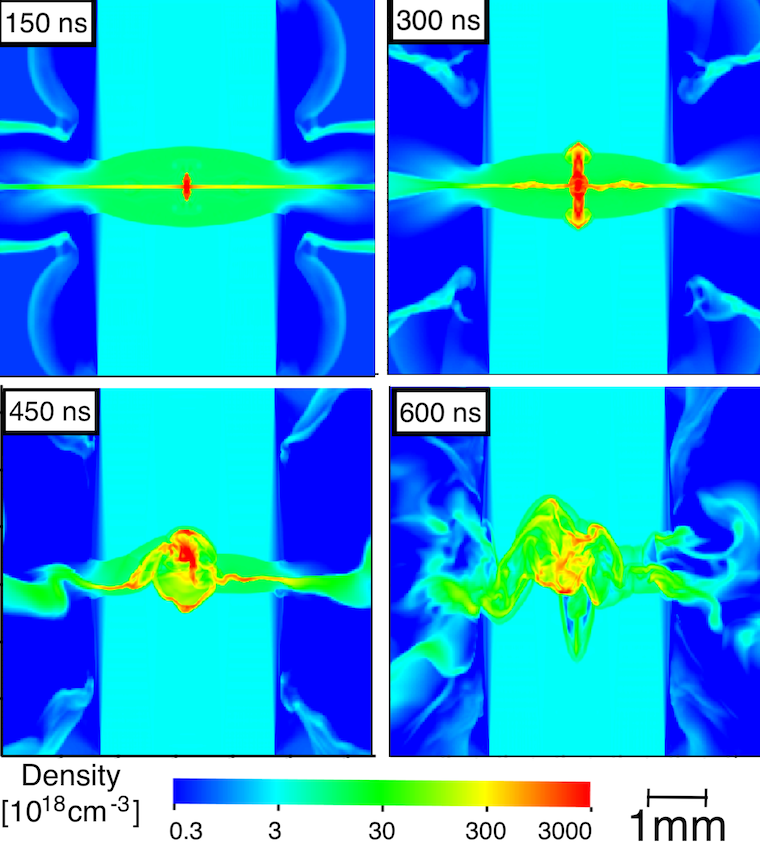}
    \caption{Run 4 from table \ref{table:3DRuns} (weak magnetic field with weak cooling which strenghtens as temperature decreases). Once again, instability of the cold slab contributes to an earlier collapse of the spine.}
    \label{fig:runs_4}
\end{figure}

We now examine a case with  $\alpha=-1.0$ and $\beta_m = 9.84$ (figure \ref{fig:runs_4}), which combines some of the features of Run 2 and Run 3. As with the other $\alpha=-1.0$ case (Run 1), the disk is more prominent when the cooling exponent is negative. Also like the other $\alpha=-1.0$ run, the strong cooling at lower temperature drives strong NTSI in the disk which then interacts with the spine. Once again we see these instabilities leading to a collapse of the spine before the spine instability has time to fully develop. 

As with the other weaker field $\beta_m=9.84$ case (Run 2), spine growth is somewhat delayed compared to the stronger field $\beta_m=0.84$ cases (see figure \ref{fig:spine_time} in section \ref{sec:discuss}). Here however the delay is far less than Run 3,  the $\alpha=+1.0, \beta_m = 9.84$ case. The more rapid onset of spine formation is likely the result of the stronger cooling at low temperature. More disk material is available in this run to feed spine growth. Once the spine begins to form however its propagation speed is the lowest of any of our $\beta_m \leq 10$ cases (see table \ref{tab:growth} in section \ref{sec:discuss}), though it is still comparable to the other $\beta_m = 9.84$ cases.

\subsection{Strong Cooling}
For our remaining simulations, we now increase the cooling strength $\Lambda_0$ by a factor of 10, resulting in cooling of order $10^{-22}$ erg cm$^{3}$ s$^{-1}$. With stronger cooling, the radiative shock and nonlinear thin shell instabilities of the cooling region and disk contribute significantly to the long-term evolution of the system, in some cases before the spine is able to properly form.

\begin{figure}
    \centering
        \figuretitle{Low $\beta_m$, low (left) vs high (right) $\Lambda_0$, $\alpha=+1$}
        \includegraphics[width=\columnwidth]{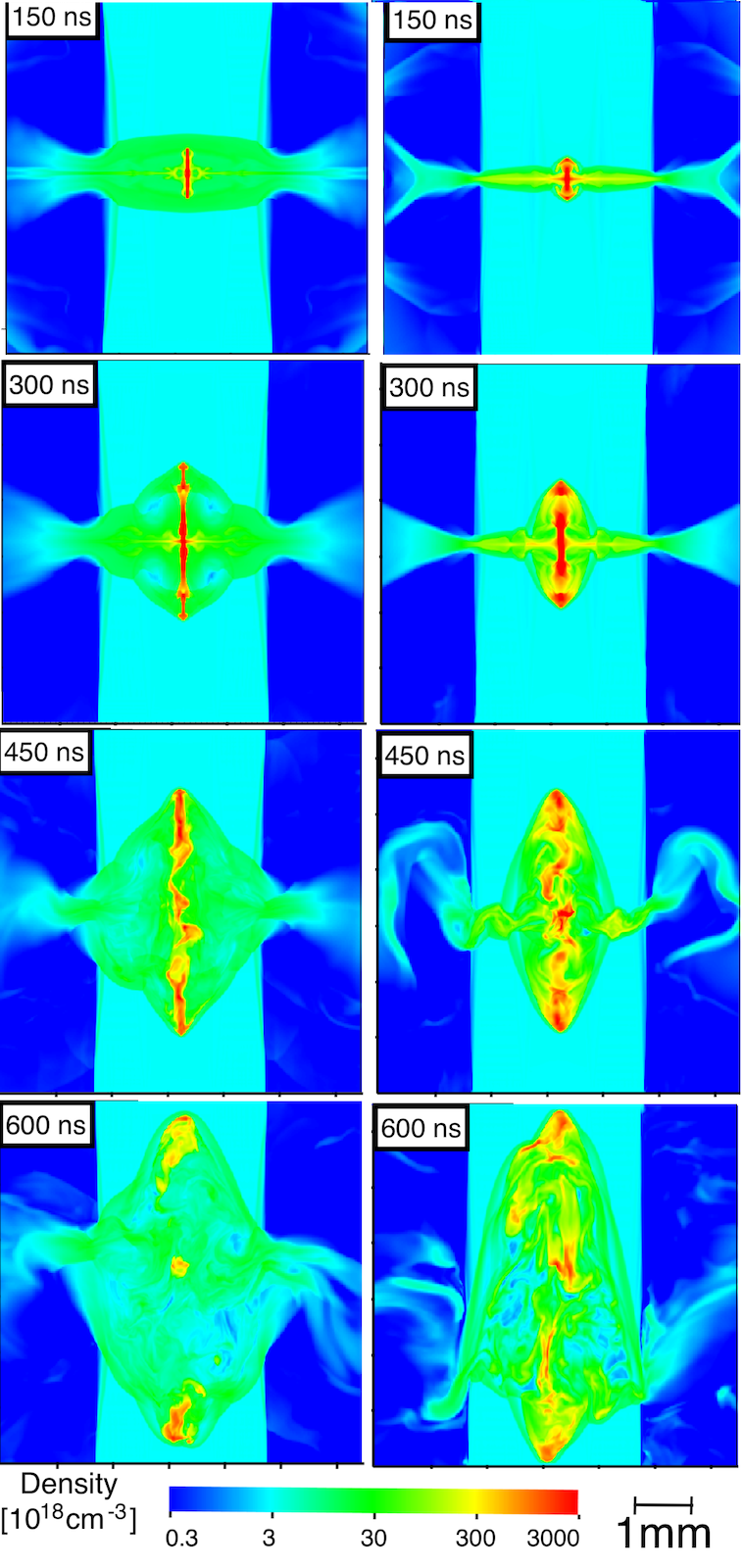}
    \caption{Runs 1 (left) and 5 (right) from table \ref{table:3DRuns}. Both feature relatively strong magnetic fields and positive $\alpha$, but the absolute strength of cooling is higher in run 5. 
    While the spines are of comparable length, stronger cooling provides more cold material, resulting in a wider spine.}
    \label{fig:runs_1_5}
\end{figure}

We begin examining the effects of this change by comparing low and high cooling strengths for the models with $\beta_m=0.84$ and $\alpha=+1.0$, (i.e. for strong field, positive cooling exponent).  Results are shown in figure \ref{fig:runs_1_5}. In the case of stronger overall cooling due to higher $\Lambda_0$, the spine emerges as earlier compared with the low $\Lambda_0$ runs.  We note that the spine has a somewhat larger radius. A prominent disk is also present. Both of these effects can be traced to the shorter cooling length expected for higher $\Lambda_0$: a smaller cooling region limits opportunity for cooling matter to escape via lateral outflows, so a greater portion of the initial flow (extending into $r>R_m$) is destined for the cold slab.

When we compare the length of the spines, we find that over a sufficient length of time cooling strength $\Lambda_0$ does not appear to have an appreciable effect. Like the $\alpha = +1.0$ cases with lower cooling, the instability of the growing spine still appears to be the dominant effect in its evolution (as opposed to any instabilities which may arise within the disk).

\begin{figure}
    \centering
        \figuretitle{Low $\beta_m$, high $\Lambda_0$, $\alpha=-1$}
        \includegraphics[width=\columnwidth]{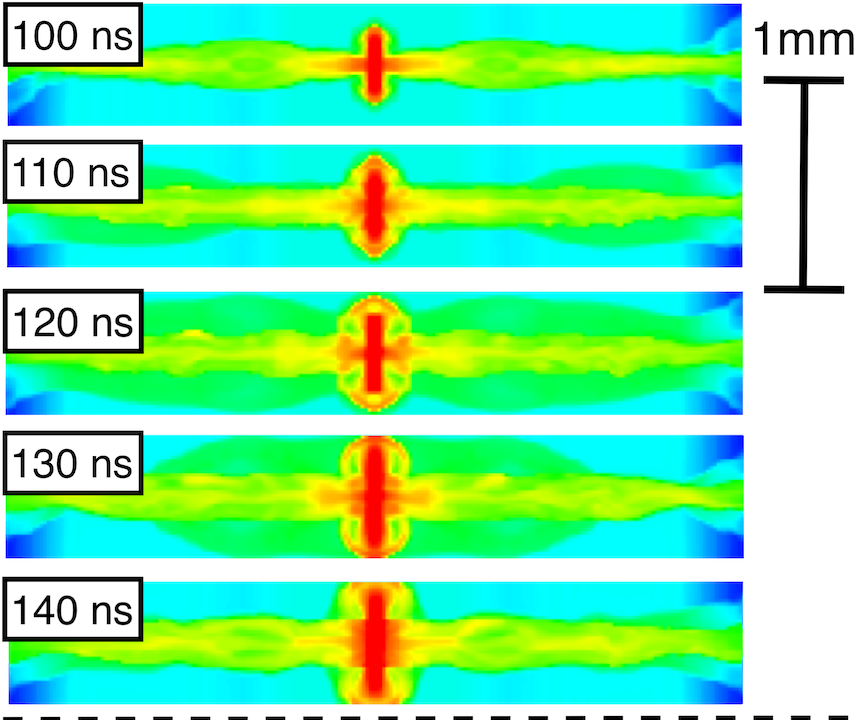}
            \includegraphics[width=\columnwidth]{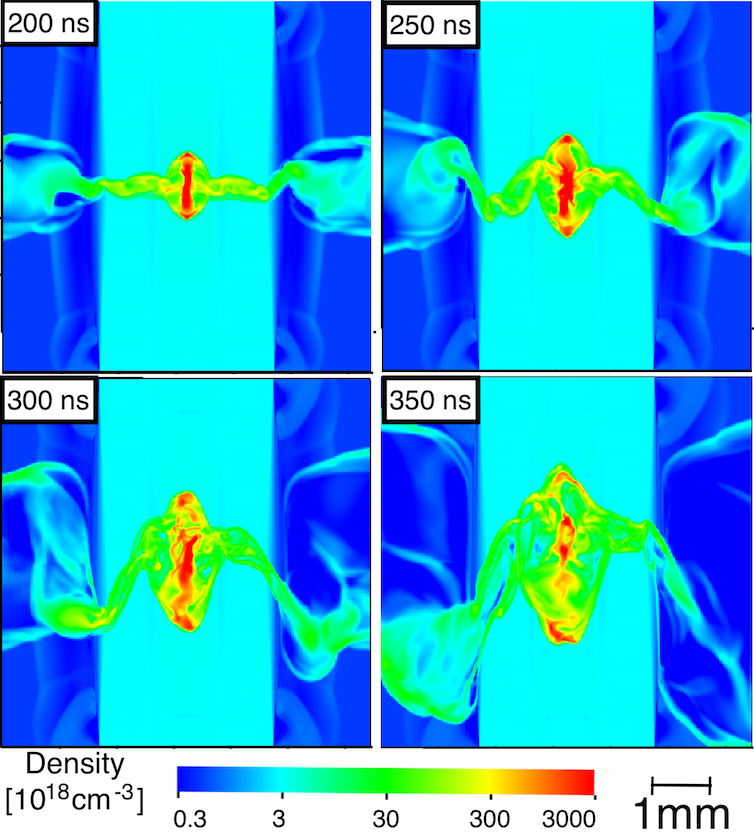}
    \caption{Runs 6 from table \ref{table:3DRuns}, which features strong fields and strong cooling which strengthens as temperature decreases.
    The first five images highlight the oscillations of the cooling region at an earlier time window. At later times we zoom out to show spine growth; stronger cooling provides more cold material, resulting in a wider spine. }
    \label{fig:run_6_osc}
\end{figure}

In figure \ref{fig:run_6_osc}, we again consider strong cooling (high $\Lambda_0$) with $\beta_m = 0.84$, but we we add the effects of $\alpha=-1.0$. 
Here we begin to see some of the hydrodynamic instabilities studied in \citet{Markwick21} play a role in shaping the evolution interaction region, though magnetic fields continue to have effects as well.

Not long after collision occurs, the radiative shock instability promotes oscillation in size of the cooling region. The timescale of these oscillations is inversely proportional to the cooling length, making them more prominent in the cases with stronger cooling.  As shown in the first five panels of figure \ref{fig:run_6_osc} and discussed in \citet{Markwick21}, these oscillations imprint variation within the disk, allowing instabilities which disrupt the entire interaction region to occur earlier than in the corresponding low $\Lambda_0$ case.

\begin{figure}
    \centering
            \figuretitle{High $\beta_m$, high $\Lambda_0$, $\alpha=+1$}
        \includegraphics[width=\columnwidth]{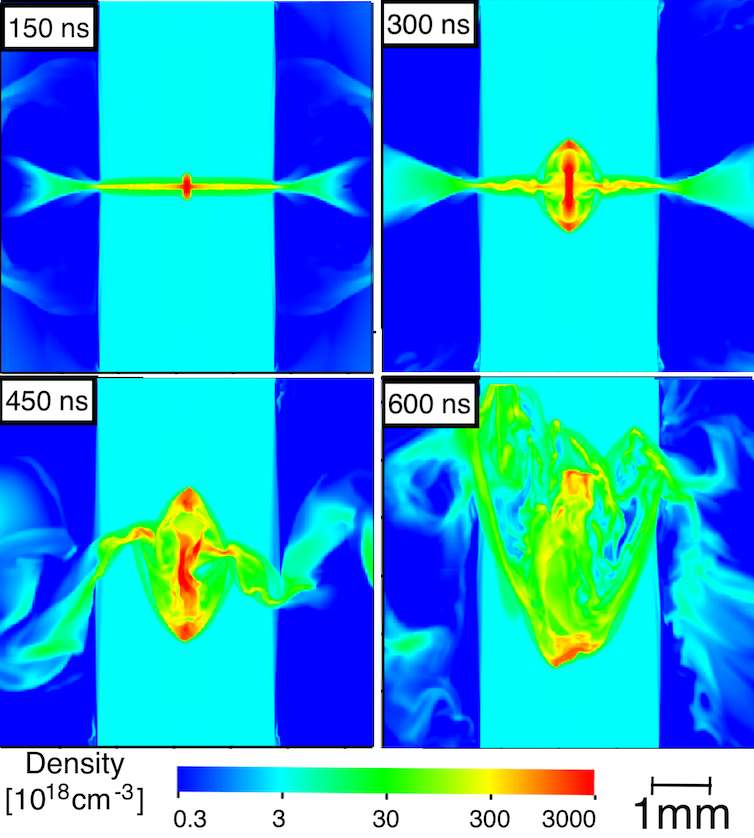}
    \caption{Run 7 from table \ref{table:3DRuns}, which features weaker magnetic fields and strong cooling which weakens as temperature decreases. Stronger cooling in the disk allows the spine to grow larger, though instabilities in the disk also contribute to the spine's eventual collapse.}
    \label{fig:runs_3_7}
\end{figure}

%We now proceed to  is that with $\beta_m = 9.84$
We now consider pairing strong cooling (high $\Lambda_0$) with weaker fields ($\beta_m=9.84$). If $\alpha=+1.0$ (figure \ref{fig:runs_3_7}), the result is an interaction region which exhibits similarities to several other cases before being disrupted by multiple simultaneous instabilities.

The disk in this case is similar to that of the corresponding $\beta_m=0.84$ case with the same cooling parameters. We see the disk is thickened by the stronger total cooling while lacking the disturbances of the radiative shock instability. The spine is also of a similar thickness to the other strong cooling cases. Finally, the spine growth rate is comparable to other $\beta_m = 9.84$ cases, but the onset of its growth is closer to the $\alpha=-1.0$, low $\Lambda_0$ case than it is to the $\alpha=+1.0$, low $\Lambda_0$ case.

As expected, with stronger overall cooling, bending modes \revision{characteristic of the nonlinear thin shell instability} appear within the disk before the spine begins to develop instabilities. As these bending modes grow, the NTSI alters disk at a similar time to when the spine undergoes its own instability.

\begin{figure}
    \centering
            \figuretitle{High $\beta_m$, high $\Lambda_0$, $\alpha=-1$}
        \includegraphics[width=\columnwidth]{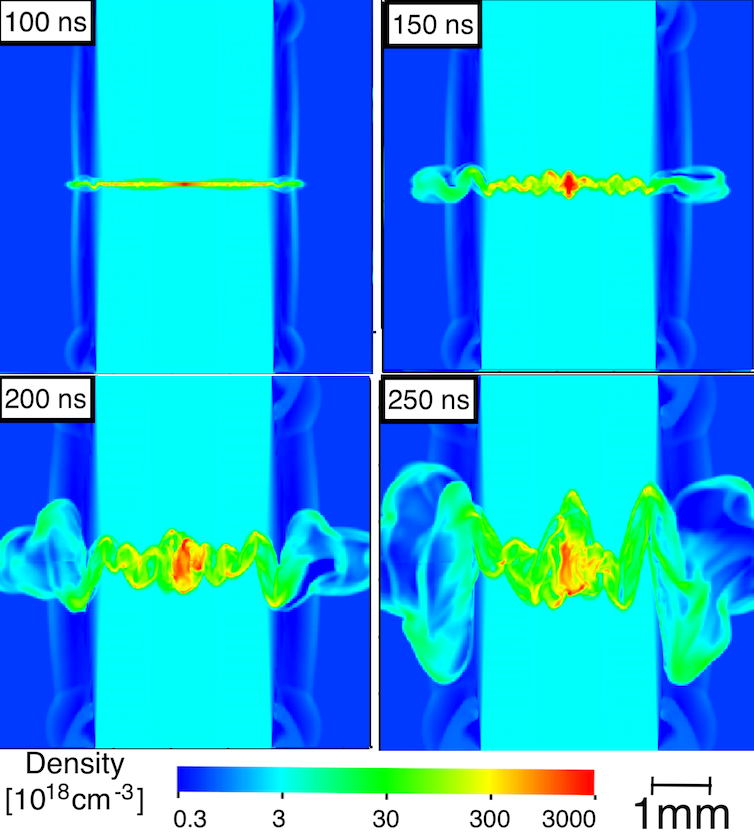}
    \caption{Run 8 from table \ref{table:3DRuns}, which features which features weaker magnetic fields and strong cooling which strenghtens as temperature decreases. The combination of radiative shock and nonlinear thin shell instability serve as the primary drivers of this run's evolution, the presence of a nontrivial magnetic field still results in cold slab material moving towards the centre.}
    \label{fig:run8}
\end{figure}

%){\bf AF. I'm not sure what this paragraph is refering too.  You never reference figure 9.  Can you try and unpack what is happening there better.  It never creates a spine.  Why?  Its a high Lambda run}) 
If we combine weaker cooling, $\alpha=-1.0$, and $\beta_m=9.84$ (shown in figure \ref{fig:run8}), the hydrodynamic instabilities occur at an earlier time and prevent the formation of a spine. As was the case in \citet{Markwick21}, the radiative shock instability imprints variation onto the cold slab, allowing the NTSI to begin at an earlier timescale than it would otherwise. 
At the same time, weaker magnetic fields ($\beta_m = 9.84$) result in a delay in the formation of the spine (We do note however that the magnetic fields do still cause cold slab material to congregate near the axis). The combination of these effects places spine formation at a later time than the disruption of the disk by the NTSI, therefore preventing the former altogether.

%\subsection{Hydrodynamic Limit}

\begin{figure}
    \centering
            \figuretitle{Hydrodynamic limit with high $\Lambda_0$}
        \includegraphics[width=\columnwidth]{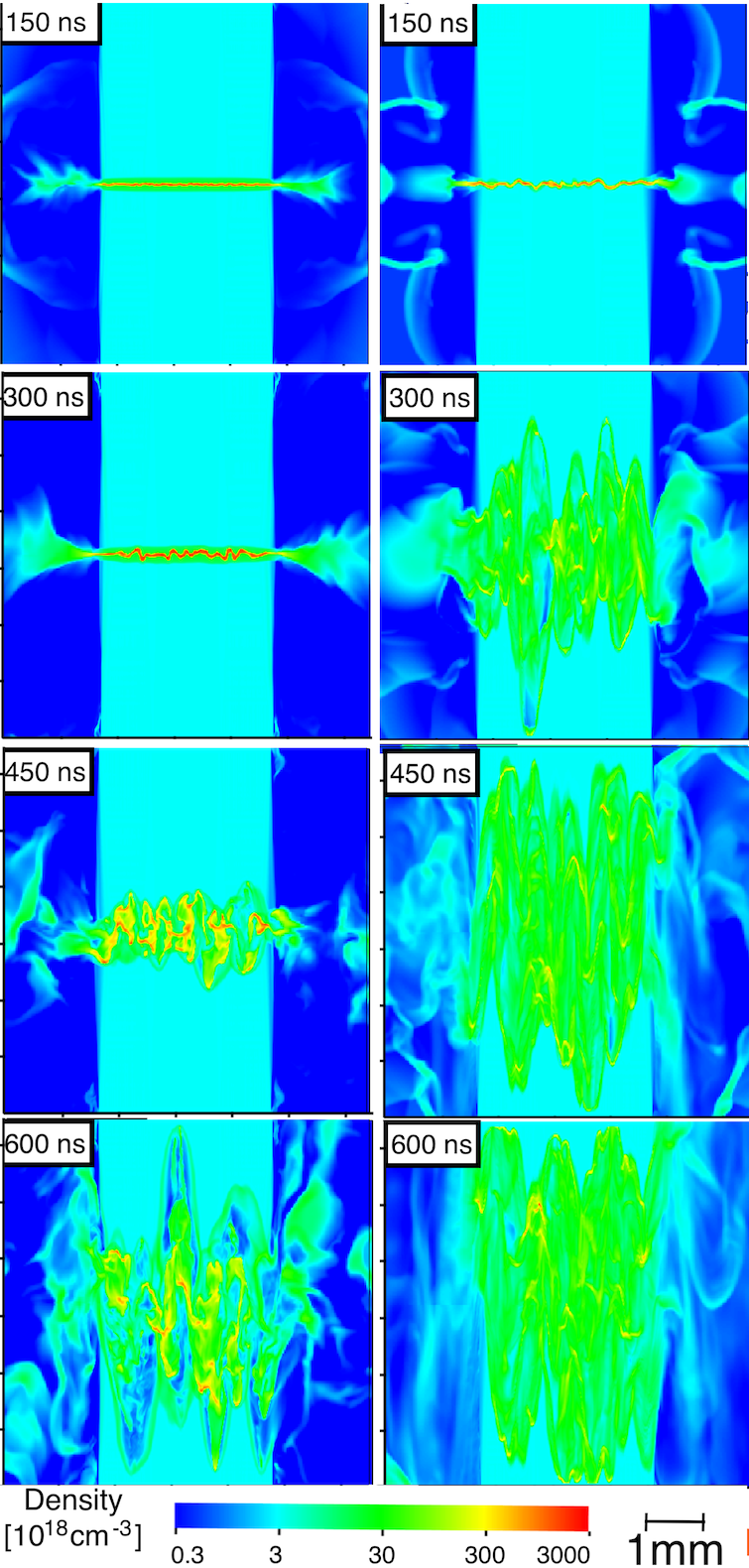}
    \caption{Runs 9 (left) and 10 (right)from table \ref{table:3DRuns}. These runs reproduce the NTSI-driven hydrodynamic case seen in \citet{Markwick21}, with the radiative shock instability enhancing the NTSI in run 10}
    \label{fig:runs_9_10}
\end{figure}

Lastly, we ran a pair of cases with $\beta_m = 10^5$, again using strong cooling. This should approximate the hydrodynamic limit we studied in \citet{Markwick21}. As seen in figure \ref{fig:runs_9_10}, we do recover the results from that paper. Without a significant magnetic field, a spine does not form. Instead the cold slab forms a disk which, over time, becomes unstable. It is the NTSI which drives bending modes which eventually disrupt the interaction regions. As expected, this occurs sooner in the case where the cooling law exponent is negative, since the radiative shock instability imprints a seed for the NTSI.

\section{Discussion} \label{sec:discuss}
\subsection{Magnetic fields in the Cold Slab}
We now present a simplified model for the formation and propagation of the spines seen in our simulations. We do this by considering unperturbed conditions for the spine formation. We must first find expressions for the density $\rho_d$, pressure $p_d$, and magnetic field $B_d$ of the disk in terms of the corresponding preshock values $\rho_i$, $p_i$, $B_i$, and jet velocity $v_i$. Also relevant to this calculation is sound speed $c_s = \sqrt{\frac{\gamma p_i}{\rho_i}}$.
Radial variation of the preshock magnetic fields will result in some of these quantities being dependent on radius. 

 Approximating the cooling region as a stationary isothermal shock \revision{(i.e. the quantity $\frac{p}{\rho}$ returns to its preshock value when it reaches the cold slab)}, 
conservation of mass and momentum can be expressed as
\begin{equation}
\rho_i\left(v_i^2 + \frac{c_s^2}{\gamma}+ \frac{B_i^2}{8\pi \rho_i}\right) = \rho_d\left( \left[\frac{\rho_iv_i}{\rho_d}\right]^2 + \frac{c_s^2}{\gamma} + \frac{B_i^2}{8\pi \rho_i} \frac{\rho_d}{\rho_i} \right)\end{equation}
This equation is cubic in $\frac{\rho_d}{\rho_i}$; solving yields
\begin{equation} \label{eq:density_jump_exact}
\frac{\rho_d}{\rho_i}= \frac{\gamma \beta_r M}{1+\beta_r} \left(\frac1{2M} + \sqrt{\frac1{4M^2} + \frac{\gamma\beta_r}{(1+\beta_r)^2} } \right)^{-1}
\end{equation}
where $\beta_r$ is value of $\frac{8\pi p_i}{B_i^2}$ at a radius $r$.
In the $\beta_r >>\gamma M^2$ limit, 
we recover the $\rho_d = \gamma M^2 \rho_i$ result found in \citet{Markwick22}. 
For $\beta << \gamma M^2$ on the other hand, we find 
\begin{equation}\label{eq:density_jump}
\rho_d(r)= M\sqrt{\gamma\beta_r} \rho_i
\end{equation}

We note that $M$ varies as $p_i^{-\frac12}$ and $\beta_m$ varies as $p_i$, so equation \ref{eq:density_jump} varies inversely with $B$ and is independent of $p_i$. 
Using equation \ref{eq:B_initial} for our initial $B(r)$, the density  and pressure jumps are therefore given as
\begin{equation}
\rho_d(r)= M\sqrt{\gamma\beta_m} \rho_i \begin{cases}\frac{R_m}{r} & 0\leq r \leq R_m \\ \frac{r}{R_m} & R_m<r \leq R_j\end{cases}
\end{equation}
\begin{equation}
p_d(r)= M\sqrt{\gamma\beta_m} p_i(r) \begin{cases}\frac{R_m}{r} & 0\leq r \leq R_m \\ \frac{r}{R_m} & R_m<r \leq R_j\end{cases}
\end{equation} 
where $p_i(r)$ is given by equation \ref{eq:p_initial}.
Flux freezing tells us that the magnetic fields grow proportional to density; the magnetic fields in the disk are approximately given as $B_d = M\sqrt{\gamma\beta_m} B_m$, which is independent of position.

\begin{figure}
    \centering
    \includegraphics[width=\columnwidth]{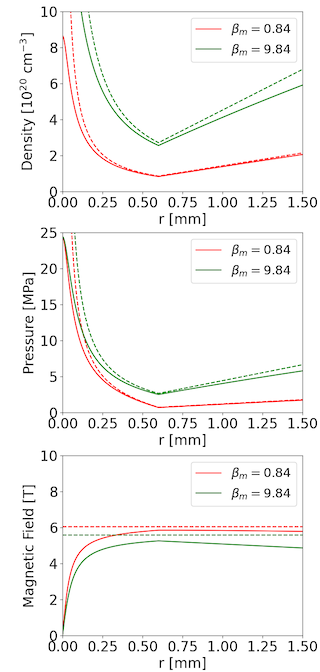}
    \caption{Estimated Density, Pressure, and Magnetic Field in the disk. The solid lines are plotted using equation \ref{eq:density_jump_exact}, while the dashed lines show the $\beta<<\gamma M^2$ approximation (equation \ref{eq:density_jump}). Away from the jet axis, the two approximations agree reasonably well, especially for lower $\beta_m$.}
    \label{fig:disk_profile}
\end{figure}

 Figure \ref{fig:disk_profile} shows density, pressure, and magnetic fields in the disk, both using the $\beta_m <<\gamma M^2$ approximation (equation \ref{eq:density_jump}) and the full isothermal approximation (equation \ref{eq:density_jump_exact}).
We find that the two approximations are reasonably close to each other, especially for stronger magnetic fields (lower $\beta_m$). The $\beta_m <<\gamma M^2$ approximation however is inaccurate at locations very close to the axis, since the magnetic fields are weak and we instead approach the hydrodynamic jump conditions.

\subsection{Growth of the spine}

 We now develop a simple model for the growth of the spine in which magnetic forces in the disk draw material towards the axis.  The build-up of magnetic and gas pressure there then create a collimated bipolar flow pushing back into both jets. 
 
 We begin by assuming the spine has uniform density $\rho_s$ and grows with velocity $v_s$. Let $2h_s$ and $r_s$ be the height\footnote{The factor of two in the height accounts for the spine growing from both sides of the disk} and radius of the spine.
 The growth of the spine is driven by the high magnetic pressure of the cold slab, so we define the total pressure of the disk to be $p_d + \frac{B_d^2}{8\pi}$. It is also fed momentum flux $-\rho_i v_i^2$ from above.  We have assumed the ram pressure from the disk itself to be negligible, as is the preshock pressure\footnote{which is of order $M^2$ less than the ram pressure} and magnetic field\footnote{which vanishes completely at the axis}.

\begin{figure}
    \centering
    \includegraphics[width=0.95\columnwidth]{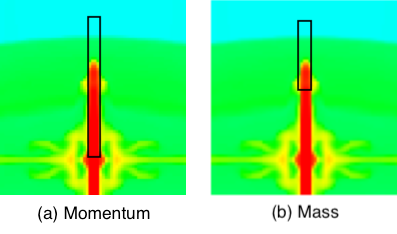}
    \caption{A diagram showing the regions for which mass and momentum conservation are to be considered for the spine.}
    \label{fig:spineRegion}
\end{figure}

 Consider a cylindrical region of radius $r_s$ and height $h_r$ centered on the spine's axis. Let one end of the cylinder be located in the preshock region, while the other end is at the disk. Such a region is pictured in figure \ref{fig:spineRegion}. 
The momentum of the spine within this region is given by $(\rho_sv_s)(\pi r_s^2 h_s)$, while the momentum in the rest of the region is given by $(-\rho_iv_i)(\pi r_s^2[h_r-h_s])$. Here we denote $v_i$ as being positive towards the disk, and $v_s$ being positive away from the disk. 
Since $\rho_i$, $v_i$, $r_s$, and $h_r$ are all constant, vertical momentum conservation is given by 
\begin{equation}
\frac{\partial}{\partial t}\left[(\rho_s v_s + \rho_iv_i) h_s\ \right]
=
\left(p_d + \frac{B_d^2}{8\pi}\right) - \left(\rho_i v_i^2\right)
\end{equation}
Assuming density and velocity to be constant, we have
\begin{equation} \label{eq:mom_balance_spine}
(\rho_s v_s + \rho_iv_i)\frac{\partial h_s}{\partial t}
=
\left(p_d + \frac{B_d^2}{8\pi}\right) - \left(\rho_i v_i^2\right)
\end{equation}

Consider now a cylinder similar to the one we have been examining, but with the lower boundary located at some distance $h_-$ above the disk (see figure  \ref{fig:spineRegion}). Mass conservation for this new cylinder is given by
\begin{equation}
\frac{\partial}{\partial t}\left[(\rho_s- \rho_c) (h_s-h_-) \right]
=\left(\rho_sv_s\right) + \left(\rho_i v_i\right)
\end{equation}
where $\rho_c$ is the density of the cooling region immediately ahead of the spine head\footnote{or the preshock density if the spine has already passed the jet shock}.
We can solve this equation for $\frac{\partial h_s}{\partial t}$:
\begin{equation} \label{eq:mass_balance}
\frac{\partial h_s}{\partial t}
= \frac{\rho_sv_s + \rho_i v_i}{\rho_s- \rho_c}
\end{equation}
Plugging this into equation \ref{eq:mom_balance_spine} gives us 
\begin{equation} %\label{eq:mom_balance_spine2}
\frac{\left(\rho_sv_s + \rho_iv_i\right)^2}{\rho_s- \rho_c}
=
\left(p_d + \frac{B_d^2}{8\pi}\right) - \left(\rho_i v_i^2\right)
\end{equation}
Assuming $\rho_c<<\rho_s$ gives us
\begin{equation} \label{eq:mom_balance_spine2}
\rho_sv_s^2
=
p_d + \frac{B_d^2}{8\pi} - \left[1+\frac{\rho_i}{\rho_s} + \frac{2v_s}{v_i}\right]\rho_i v_i^2
\end{equation}
Since $\rho_i\leq \rho_c$ we must also have $\rho_i<<\rho_s$, and if we assume\footnote{This assumption could be made on grounds of mass conservation, or the assumption that spine growth is fed equally from above and from below} $\rho_s v_s \sim \rho_i v_i$ we can also infer $v_s<<v_i$. We therefore can eliminate the factor in square brackets as being close to unity:
\begin{equation}
\rho_sv_s^2 = p_d + \frac{B_d^2}{8\pi} - \rho_i v_i^2 
\end{equation}

Let us strengthen our assumption to  $\rho_s v_s = \rho_i v_i$ exactly. Dividing by $\rho_s$ gives us
\begin{equation}
v_s^2 = \frac{p_d}{\rho_s} + \frac{B_d^2}{8\pi\rho_s} -  v_sv_i 
\end{equation}
Solving the quadratic gives
\begin{equation}
v_s = \frac12\left(v_i^2 + \frac{4p_d}{\rho_s} + \frac{B_d^2}{2\pi\rho_s}\right)^{\frac12} - \frac{v_i}{2}  
\end{equation}
which can be approximated as
\begin{equation}
v_s = \frac{p_d}{\rho_s v_i} + \frac{B_d^2}{8\pi\rho_s v_i} 
\end{equation}
Given that the cooling region is approximately an isothermal shock, $\frac{p_d}{\rho_s}$ is equal to $\frac{p_i}{\rho_i}$, which is in turn $\frac{v_i^2}{\gamma M^2}$ by definition.
\begin{equation}
v_s = \frac{v_i}{\gamma M^2} + \frac{B_d^2}{8\pi\rho_s v_i} 
\end{equation}

To replace the remaining terms $B_d$ and $\rho_s$ with their preshock counterparts, we consider how alfven speed $v_{A} = \frac{B}{\sqrt{4\pi\rho}}$ scales across a shock. The density increases by a factor of $M\sqrt{\gamma\beta_m}$, but so does the magnetic field as a result of flux freezing. The alfven speed therefore increases by a net factor of $M^{\frac12} (\gamma\beta_m)^{\frac14}$, leaving our overall equation as 
\begin{equation}
v_s = \frac{v_i}{\gamma M^2} + M(\gamma\beta_m)^{\frac12} \frac{B_m^2}{8\pi\rho_i v_i} 
\end{equation}
Since $M\sqrt{\gamma\beta_m} = v_i\sqrt{\frac{8\pi \rho_i}{B_m^2}}$, this is equal to 
\begin{equation}
v_s = \frac{v_i}{\gamma M^2} +  \frac{B_m}{\sqrt{8\pi\rho_i}} 
\end{equation}
Finally, going back to equation \ref{eq:mass_balance} tells us that the growth speed is approximately twice this, or
\begin{equation} \label{eq:growth_speed}
\frac{\partial h_s}{\partial t}
= \frac{2v_i}{\gamma M^2} +  \frac{B_m}{\sqrt{2\pi\rho_i}}
\end{equation}.
In terms of $\beta_m$, this can be expressed as
\begin{equation}
\frac{\partial h_s}{\partial t} = 2v_i \left(\frac{1}{\gamma M^2} +  \frac{1}{M\sqrt{\gamma\beta_m}} \right)
\end{equation}

\begin{figure}
    \centering
    \includegraphics[width=\columnwidth]{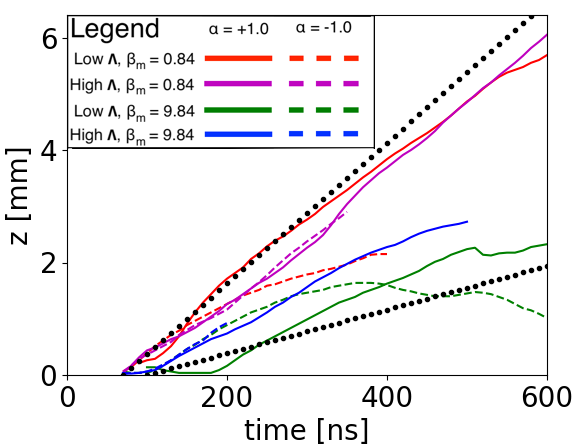}
        \includegraphics[width=\columnwidth]{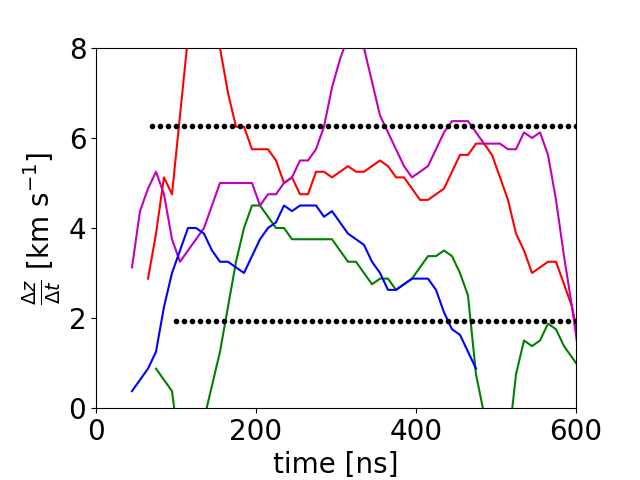}
    \caption{A) plot of the size of the spine vs time.
    B) a time derivative of the spine size in the $\alpha=+1.0$ cases, averaged over intervals of 50ns. 
    In both plots, the dotted black lines match the estimated growth speeds for $\beta_m = 0.84$ (upper line) and $\beta_m = 9.84$ (lower line).    }
    \label{fig:spine_time}
\end{figure}

To compare our estimate to our simulations, we selected a starting and stopping time for each simulation and measured the change in $h_s$ over that time interval. The results are shown in table \ref{tab:growth}, and a time plot of spine growth is shown in  figure \ref{fig:spine_time}. 

We find that the model predicts growth speeds which compare well with the simulations given the model's simplicity, with differences ranging between approximately $10\%$ to 50\%

For runs 1 and 5 (which feature $\beta_m = 0.94$ with $\alpha=-1$), we find that the model predicts growth speeds which overestimate the speeds by only about 12\%.  Runs 2 and 6, which are the corresponding $\alpha=-1$ cases, are slower than their $\alpha=+1$ counterparts despite this not being accounted for in the estimation, while the $\beta_m=9.84$ cases (runs 3, 4, 7, and 8) show faster observed growth than is estimated.  In general, runs with positive cooling exponent and lower beta (stronger field) produce the best correspondence to the model.

% It may be worth noting that introducing a factor of $\beta_m^{\frac14}$ would improve the agreement for the $\beta_m=9.84$ cases to a margin which exceeds that of  the $\beta_m=0.84$ cases (which would also see a small improvement), though it is unclear how such a factor might arise.

\begin{table}
    \centering
    \begin{tabular}{|c||c|c|c|c||c|c|}\hline
    Run & $\beta_m$ & $t^\text{start}$& $t^\text{stop}$& $\Delta h_s$ & $\frac{\Delta h_s}{\Delta t}$ & $\left(\frac{\partial h_s}{\partial t}\right)_\text{est}$\\\hline\hline
1 & 0.84 & 90 & 550 & 2.58 & 5.61 & 6.26  \\\hline
2 & 0.84 & 70 & 390 & 1.04 & 3.26 & 6.26  \\\hline
3 & 9.84 & 180 & 510 & 1.11 & 3.37 & 1.94  \\\hline
4 & 9.84 & 100 & 360 & 0.80 & 3.07 & 1.94  \\\hline
5 & 0.84 & 70 & 550 & 2.70 & 5.62 & 6.26  \\\hline
6 & 0.84 & 70 & 350 & 1.41 & 5.06 & 6.26  \\\hline
7 & 9.84 & 80 & 500 & 1.35 & 3.21 & 1.94  \\\hline
8 & 9.84 & 70 & 200 & 0.45 & 3.50 & 1.94  \\\hline
    \end{tabular}
    \caption{The observed speeds for the growth of spine size $h_s$ in the simulations, compared to the estimated speeds for the corresponding value of $\beta_m$. Times are given in ns, $\Delta h_s$ is given in mm, and speeds are given in km s$^{-1}$. }
    \label{tab:growth}
\end{table}

\subsection{Timescale Analysis of Disk Instabilities}

In addition to unstable behaviour which causes the spine to break apart, there are two additional hydrodynamic instabilities occurring within the interaction region: the NTSI in the disk, and the radiative shock instability in the cooling region.  The latter affects the system by imprinting variation onto the disk and promoting the NTSI.  

Before examining these two instabilities further, we briefly comment on the breakup of the spine. The nature of the dynamics in this case, in terms of a dominant mode, is unclear as the $\beta$ value in the spine is intermediate between hydrodynamic (in which instabilities such as Kelvin Helmholtz might apply) and magnetically dominated (in which instabilities such as kink or sausage might apply) regimes. We observe timescales for the spine to break-up on the order of a few hundred nanoseconds after the onset of its growth. We leave a more detailed analysis of its dynamics for a future project and focus on the the jet interaction region.

\subsubsection{Nonlinear Thin Shell Instability}
The nonlinear thin shell instability promotes growth of bending modes within the disk. We define the amplitude of a perturbation as $\psi$\revision{, and its wavelength as $\lambda$}. Its growth speed in the hydrodynamic case given as
\begin{equation} \label{eq:NTSI_growth}
\dot \psi \sim \left(\frac{\psi}{\lambda}\right)^{\frac32} c_s.
\end{equation}
\citep{Vishniac94}. 
 An appropriate timescale for growth would therefore be
\begin{equation}
\tau_\text{NTSI} \sim \frac{\psi}{\dot\psi} \sim \left(\frac{\lambda}{\psi}\right)^{\frac12} \frac{\lambda}{c_s}.
\end{equation}

There is a constraint that $\psi$ and $\lambda$ must both exceed the thickness of the system (i.e. of order $d_\text{cool}$). With this in mind, we approximate our initial timescale as
\begin{equation} \label{eq:NTSI}
\tau_\text{NTSI}\sim M \frac{d_\text{cool}}{v_i}.
\end{equation}
We therefore find that for our strong cooling cases we have an estimated timescale of order 170 ns; this appears to be in agreement with the evolution of the $\alpha=+1.0$ hydrodynamic case presented in figure \ref{fig:runs_9_10}. 

We must consider the impact of magnetic fields on the NTSI. \citet{Heitsch} found that magnetic fields perpendicular to the shock can inhibit the NTSI, but that the instability will often still develop. 
For toroidal magnetic fields, we expect magnetic tension to only resists perturbations with azimuthal variation; radial variation would be perpendicular to the magnetic field, yielding magnetic tension  equal to the unperturbed case.

Assuming a purely azimuthal perturbation of the form $\psi \revision{\sim e^{im\phi}}$ and a sufficient distance from the axis, the Euler equation can be expressed as 
\begin{equation} \label{eq:NTSI_Euler}
\rho \ddot \psi \sim  \left(\rho \ddot \psi \right)_\text{hydro} +   \frac{B_\phi}{4\pi r}\frac{\partial B_z}{\partial \phi}  \end{equation}
where $\dot\psi_\text{hydro}$ is the growth in velocity  the hydrodynamic case, which per equation \ref{eq:NTSI_growth} is governed by
\begin{equation}
 \ddot \psi_{\revision{\text{hydro}}}
\sim \frac32\left(\frac{\psi}{\lambda}\right)^{\frac12} \frac{\dot\psi }{\lambda}c_s \sim \frac32\left(\frac{\psi}{\lambda}\right)^{2} \frac{c_s^2}{\lambda}
\end{equation}. 
Meanwhile flux freezing should bend the field into a sinusoidal shape to match our perturbation, giving us $B_z$ of order $\frac{B_\phi}{r}\frac{\partial \psi}{\partial\phi}$.
Putting these two results along with $\lambda = \frac{2\pi r}{m}$ into equation \ref{eq:NTSI_Euler} gives us
\begin{equation} 
 \ddot \psi \sim  \frac32\left(\frac{\psi}{\lambda}\right)^{2} \frac{c_s^2}{\lambda} -\frac{4\pi^2 a^2}{\lambda^2} \psi  \end{equation}
where $a = \frac{B_\phi}{\sqrt{4\pi\rho}} $ is the local alvfen speed.
Let $\beta_m' = \frac{2c_s^2}{\gamma a^2}$  be the local plasma beta; we can express our growth law as
\begin{equation}
 \ddot \psi \sim   \frac32\left(\frac{\psi}{\lambda}\right)^{2} \frac{c_s^2}{\lambda} \left[1 - \frac{16\pi^2}{3 \gamma \beta_m'} \left(\frac{\lambda}{\psi}\right) \right]
\end{equation}
Since $\beta_m'$ in the disk is smaller than global $\beta_m$ by a factor of order $M\sqrt{\gamma\beta_m}$, we have
\begin{equation}
 \ddot \psi \sim   \frac32\left(\frac{\psi}{\lambda}\right)^{2} \frac{c_s^2}{\lambda} \left[1 -   \frac{16\pi^2}{3}\left(\frac{\gamma M^2}{\beta_m}\right)^{\frac12}\left(\frac{\lambda}{\psi}\right) \right] 
\end{equation}.

\begin{figure}
    \centering
        \figuretitle{Annular Structure for the Nonlinear Thin Shell Instability}
        \includegraphics[width=\columnwidth]{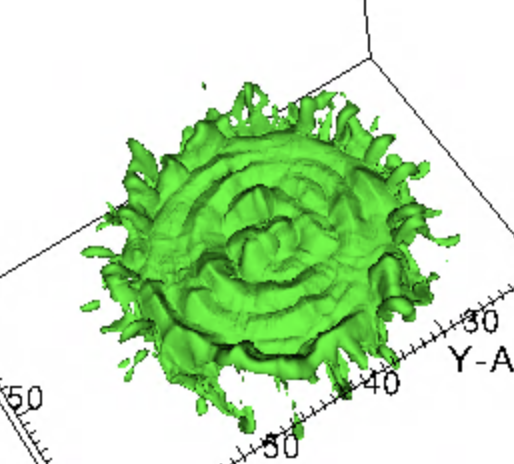}
    \caption{An isosurface plot of density in the $\alpha=-1$, strong cooling, weak field case. Note that perturbations with azimuthal variation are suppressed, with the most prominent perturbations featuring annular structure.}
    \label{fig:ntsi_ring}
\end{figure}

The above calculation suggests that magnetic tension will prevent growth of toroidal perturbations smaller than
$\psi \sim \frac{16\pi^2}{3}\left(\frac{\gamma M^2}{\beta_m}\right)^{\frac12}\lambda$, which is quite large compared to the wavelength. However, as noted earlier, radial perturbations (such as those seeded by the radiative shock instability) should be unaffected by magnetic tension. \citet{Heitsch} also found that other conditions such as sufficient turbulent energy may also allow magnetic tension to be overcome. Examining our simulations (e.g. figure \ref{fig:ntsi_ring}) we find that perturbations do in fact adopt an annular structure; the little azimuthal variation that is present appears to be most prominent along the grid axes, suggesting numerical artefact. 

\subsubsection{Radiative Shock Instability}
The radiative shock instability produces oscillations in the size of the cooling region; these oscillations can disrupt the rest of the interaction region by imprinting spatial variation on the cold slab, which excites the NTSI. An appropriate time scale for the radiative shock instability is that of its oscillation period, which is given by \citep{CI}:
\begin{equation}
\tau_\text{RSI} = f(\alpha) t_\text{cool} \approx 4f(\alpha) \frac{d_\text{cool}}{v_j}
\end{equation}
where $f(\alpha) = \frac{\pi}{2\delta_I}$ for $\delta_I$ given in table 1 of \citet{CI}. For $\alpha=-1$, $f(\alpha) = 5.97$ for the fundamental mode and $f(\alpha) =1.65$ for the first overtone. 

In figure \ref{fig:run_6_osc} shows this case undergoing an oscillation period over the course of 40 seconds or so.
In \citet{Markwick21} we found the cooling time to be approximately
\begin{equation}
t_\text{cool} =  \frac{5k T_s}{n_s\Lambda(T_s)} \left( \frac{1-\left(\frac{T_f}{T_s}\right)^{2-\alpha}}{2-\alpha}\right)   
\label{eq:tcool}
\end{equation} 
where $n_s$ is the immediate postshock density. Using this estimate, we find that the first overtone has an estimated period of 57 ns.

%Curiously, if the same calculation is performed for the hydrodynamic case presented in \citet{Markwick21}, we find that the observed period is closer to the fundamental in that case (as opposed to the first overtone in our magnetic case).

Comparing timescales for the radiative shock and nonlinear thin shell instabilities offers insight into the observed behaviour of the disk. 
At early times, the radiative shock instability -- which for high mach number has a shorter initial timescale than the NTSI --  
causes growth to occur sooner than it does as a result of the NTSI alone, as seen in figure \ref{fig:runs_9_10} and discussed in \citet{Markwick21}.
As perturbations grow, the timescale for the NTSI is reduced, so this instability has a tendency to dominate at later times if present.

\section{Comparison to Experimental Results} \label{sec:exp}

One of the primary goals of our simulations is to better understand the experiments performed by \citet{suzukiVidal15}. In those experiments, a region of compressed material is observed behind the leading bow shock. 
As seen in figure 3 of that paper, the height of the compressed region is more than twice width of the jets, while none of our previous hydrodynamic simulations  \citep{Markwick21, Markwick22} showed an interaction region which was taller than the width of the jets. 
In our present work however we see that that magnetic fields allow for the formation of a spine, which can grow to such a height. We also see that the NTSI and radiative shock instabilities continue to be present in magnetized jets; together with the instability of the spine, these instabilities provide a mechanism for the observed fragmentation of the interaction region.  Best exhibiting the combination of these effects are cases 6 (figure \ref{fig:run_6_osc}) and 7 (figure \ref{fig:runs_3_7}); both of these feature cooling which is strong enough for hydrodynamic instabilities to contribute, but not so strong (relative to the magnetic field) that the instabilities disrupt the system before a spine is able to fully form.

While our latest set of simulations has brought us closer to the experiments of  \citet{suzukiVidal15}, it is worth noting  that key differences remain. 
First, we  note that the simulations presented in this paper used an analytic cooling model, while cooling in the experiments has a more complicated dependence on temperature and density. A more accurate model was discussed in \citet{suzukiVidal15} and was used in \citet{Markwick22}.
Second, the experiments do not show a spine in the same way that our simulations do: a spine-like structure is observed on  one side of the interaction region, but unlike our simulations it only is observed one one side of the cold slab. This may be a result of asymmetry: the experiments feature jets of differing speeds and densities.
We previously studied the effects of collisions between non-identical jets in \citet{Markwick22}, but those simulations did not include magnetic fields and thus lacked a spine.
In a future work we hope to combine the use of a more sophisticated cooling function and non-identical jets studied with the presence of magnetic fields.

\section{Conclusion} \label{sec:conc}

In this paper we have extended the results of \citet{Markwick21} by examining the structure and evolution of colliding flows with toroidal magnetic fields.
The hoop stress of the toroidal fields provides a collimating force which results in the emergence of a dense spine along the axis. The growth speed of the spine is determined primarily by the strength of the magnetic fields. 
After some time, one or more instabilities will contribute to the disruption of the interaction region. For flows with sufficient magnetic fields for spine formation, the spine eventually becomes unstable and collapses after growing for a sufficient amount of time. For flows dominated by radiative cooling, the hydrodynamic instabilities studied in \citet{Markwick21} will disrupt the cold slab, which in turn disrupts the spine at an earlier time than would occur as a result of the spine instability alone.

The results of colliding flow experiments (both those performed in the laboratory as well as those performed via simulation) can be used to gain insight into astronomical phenomena such as YSO jets.  While astrophysical jets do not ordinary collide in the same manner as in experiments, the formation of shocks produced by flow collisions can be connected (via reference frame transformation) to those formed by jet pulsation in which a slower region is overtaken by a faster region of the jet behind it \citep[e.g.][]{Gardiner00}. 
Spine growth can be observed in simulations such as \citet{Hansen15}, but in that case the axisymmetric nature of the simulation may have inhibited instabilities; our work here offers some insight into the lateral structure of the spine and its eventual breakup.
Conversely, the 3D simulations of \citet{Hansen17} exhibit the nonlinear thin shell instability, but do not exhibit spine growth or instability as they lack magnetic fields.  Real astrophysical jets are produced by magnetic effects \citep[e.g.][]{BlandfordPayne, LyndenBell96}, and so the interaction of magnetic fields with  instabilities must be considered.

\section{Acknowledgments}

This work used the computational and visualization resources in the Center for Integrated Research Computing (CIRC) at the University of Rochester. Financial support for this project was provided by the Department of Energy grants GR523126, DE-SC0001063, and DE-SC0020432, the National Science Foundation grant GR506177, and the Space Telescope Science Institute grant GR528562.  E.G. Blackman acknowledges working in part  at the Aspen Center for Physics, which is supported by National Science Foundation grant PHY-2210452. 

\section{Data Availability}

The data underlying this article will be shared on reasonable request to the corresponding author.

\bibliography{paper.bib}

\bsp

\label{lastpage}

\end{document}